\documentstyle[12pt]{article}
\def\beq{\begin{equation}}
\def\eeq{\end{equation}}
\def\tb{{\bar t}}
\def\bb{{\bar b}}
\def\msb{{\overline {MS}}}
\def\as{{\alpha_S}}
\begin{document}

\begin{titlepage}
\begin{center}
{\Large \bf Theoretical Physics Institute \\
University of Minnesota \\}  \end{center}
\vspace{0.1in}
\begin{flushright}
UMN-TH-1241/94 \\
TPI-MINN-94/5-T \\
January 1994
\end{flushright}
\vspace{0.1in}
\bigskip
\begin{center}
{\Large \bf On normalization of QCD effects in $O(m_t^2)$
electroweak corrections\\}
\vspace{0.1in}
{\bf Brian H. Smith \\}
School of
Physics \& Astronomy, University of Minnesota, Minneapolis, MN 55455 \\ and
\\
{\bf M.B. Voloshin  \\ } Theoretical Physics Institute, University of
Minnesota, Minneapolis, MN 55455 \\ and \\ Institute of Theoretical and
Experimental Physics, Moscow 117259 \\
\vspace{0.2in}
{\bf   Abstract  \\ }
\end{center}
We point out that, contrary to some recent claims, there is no intrinsic
long-distance uncertainty in perturbative calculation of the QCD
effects in the $t \tb$ and $t \bb$ loops giving the electroweak corrections
proportional to $m_t^2$. If these corrections are expressed in terms of
the ``on-shell" mass $m_t$, the only ambiguity arising is that associated
with the definition of the ``on-shell" mass of a quark. The latter is
entirely eliminated if the result is expressed in terms of $m_t$ defined at
short distances. Applying the Brodsky-Lepage-Mackenzie criterion for
determining the natural scale for normalization of $\as$, we find that
using the ``on-shell" mass makes this scale numerically small in units of
$m_t$. Specifically, we find that by this criterion the first QCD correction
to the $O(m_t^2)$ terms is determined by $\as^\msb(0.15 m_t)$.
Naturally, a full calculation of three-loop graphs is needed to completely
quantify the scale. \\
\end{titlepage}

The present remarkable statistics and accuracy of the LEP data at the energy
of the $Z$ peak calls for a significant theoretical precision of calculation
of the electroweak loop effects with the goal of sensing the effects of
Higgs boson and/or new physics once the top quark mass is known (for a
recent review see e.g.  Ref. \cite{nov}). Due to the large mass of the top
quark, the effects of the $t \tb$ and $t \bb$ loops, proportional to $m_t^2$,
have to be calculated including the QCD corrections. Specifically, the
leading in the limit $m_t^2/m_Z^2 \gg 1$ electroweak corrections, related to
the $W$ and $Z$ propagators, are universally determined$^{\cite{nov}}$ by
the finite difference of the longitudinal parts of the $Z$ boson and the $W$
boson vacuum polarization insertions at $q^2 \to 0$. When expressed for
definiteness in terms of the correction to the electroweak $\rho$ parameter
the effect of the heavy quark loop is given by

\beq
\Delta \rho= { {3\, G_\mu} \over {8 \pi^2 \sqrt{2}}}\, m_t^2 \, \left ( 1 -
{{2 \as} \over { 9 \pi}} (\pi^2 +3) \right )
\label{rho}
\eeq
in the lowest$^{\cite{veltman}}$ and the first$^{\cite{dv}}$ orders in
$\as$. The present accuracy of the data already makes necessary a
quantitative understanding of the magnitude of the $\as$ term as well as of
the higher QCD corrections.

Recently a doubt was cast$^{\cite{ks1,ghv,ks2}}$ on the calculability of
the higher QCD effects in the expression (\ref{rho}) in terms of the QCD
coupling, normalized at distances of order $m_t^{-1}$: $\as(m_t)$.
The reason for this doubt arises in a calculation of the vacuum polarization
at $q^2 \approx 0$, i.e. far below the $t \tb$ and $t \bb$ thresholds,
using the dispersion relations, which involve integrals over the spectral
densities of the physical states, containing $t \tb$ and $t \bb$ near and
above the corresponding thresholds. The resonances and the continuum states
near the threshold are governed by the perturbative and
non-perturbative dynamics at long distances, which are much larger than
$m_t^{-1}$, hence the doubt about the calculability of the vacuum
polarization at $q^2 \to 0$ in terms of $\as(m_t)$. Here we point out that
this doubt is ungrounded and that the solution of this problem is known long
ago: at least since the development of the QCD sum rules for
charmonium$^{\cite{sr}}$.  Moreover, this is exactly the central point of
the QCD sum rules, that though each individual hadronic state is governed by
{\it long-distance} dynamics, the dispersion integrals over these states,
which give the vacuum polarization far below the threshold, are determined
by the {\it short-distance} QCD dynamics. This point, which was also
emphasized in a recent paper \cite{nov1}, is further discussed below in the
text.

The long-distance QCD effects however produce a certain effect on the
expression (\ref{rho}) through the conventions associated with it.
Namely, the result in eq.(\ref{rho}) is written in terms of the on-shell
mass $m_t$, which as discussed in this paper effectively lowers the
appropriate normalization point for $\as$ in eq.(\ref{rho}) through the
contribution of the near-mass-shell region to the evolution of the
quark mass from the mass shell to distances of order $m_t^{-1}$, which are
relevant in the loops. In connection with this observation it should be
emphasized that this effect by no means makes the perturbative calculation
uncontrollable: the notion of the on-shell quark mass is consistent in any
finite order of perturbation theory in QCD and the discussed effect
manifests itself in a numerical, rather than parametrical reduction of the
appropriate normalization momentum scale for $\as$  in units of $m_t$.
Equivalently, this implies that if the QCD corrections to the
$O(m_t^2)$ electroweak corrections are expressed in terms of $\as(m_t)$
the coefficients of higher terms should be unnaturally large.
Beyond the perturbation theory there is the known problem of defining the
on-shell quark mass, which amounts to an uncertainty of the order of
$\Lambda_{QCD}$, which effectively places the limit on the accuracy of
definition of the quark mass. This is the well known case for the $b$ quark
and this particular uncertainty should be the same for the $t$ quark.
For the latter the situation is somewhat additionally complicated by the
large width of the decay $t \to W~ b$. Nevertheless, one can, quite
probably, get to the accuracy of defining the ``on-shell" top mass better
than $\sim  1 ~GeV$.

One can notice however that the problem of properly defining $m_t$ is
somewhat artificial for the electroweak corrections at the $Z$ peak. This
problem can be completely eliminated by expressing those corrections through
the top mass, measured at short distances in some other measurable quantity,
determined by the short distance dynamics. As examples of such quantities
one can pick the $m_t$ entering the electroweak corrections to the $Z \to b
\bb$ decay rate through the $t \tb W$ triangle or the total width of the
decay $t \to W~ b$, either of which is determined by distances of order
$m_t^{-1}$, or any other measurable $m_t$-dependent quantity of the same
nature. In other words, the relation between the electroweak corrections of
the type in eq.(\ref{rho}) and the quantities like the total decay rate of
top should not contain unnaturally large or small numerical coefficients in
units of $m_t$ in the scale of normalization of $\as$.

Naturally, to completely fix the normalization point for $\as$ in
eq.(\ref{rho}) one needs a full three-loop calculation in the order $\as^2$,
which has not been done yet. However as pointed out some time ago
by Brodsky, Lepage and Mackenzie$^{\cite{blm}}$ (BLM) one gets an
appropriate estimate of the normalization point and thereby makes the
coefficient of the higher order term reasonable by evaluating the lower
order graphs with an explicitly running coupling constant. Formally, this
corresponds to tagging the dependence of the higher loop term on the number
of light quark flavors $n_f$ and then shifting this dependence in the
combination $b_0=11-{2 \over 3} n_f$ into the definition of the
normalization point of $\as$ in the lower term. Here we apply this procedure
to the calculation of the electroweak correction in eq.(\ref{rho}), and find
that the BLM criterion gives the normalization point of $\as$ in the $\msb$
scheme as low as $0.15\,m_t$. The appearance of the small coefficient $0.15$
is mainly due to the usage of the on-shell top quark mass.

For the practical calculation we use the simple fact$^{\cite{barb}}$ that
when calculating the correction in eq.(\ref{rho}) in the limit $m_Z^2/m_t^2
\ll 1$ one can neglect the masses of gauge bosons altogether, which is
equivalent to setting the electroweak gauge couplings to zero (notice that
the expression in eq.(\ref{rho}) contains as the overall factor only the top
quark Yukawa coupling $h_t$).  Therefore the quantity of interest can be
expressed in terms of the dynamics of only the scalar sector coupled to the
$t$ and $b$ quarks. In these terms the correction $\Delta \rho$ is expressed
through the vacuum polarization by the quark density operators coupled to the
Goldstone bosons $\chi^0$ and $\chi^\pm$:

\begin{eqnarray}
P_0(q^2)&=&-2i \, \int \langle 0 | T \left ( (\tb(x)\, \gamma_5 \, t(x)) \,
(\tb(x)\, \gamma_5 \, t(x)) \right ) | 0 \rangle \, e^{i q x} \, d^4 x  ~~,
\nonumber \\
P_\pm(q^2)&=&i \, \int \langle 0 | T \left ( (\tb(x)\,(1- \gamma_5) \, b(x))
\, (\bb(x)\,(1+ \gamma_5) \, t(x)) \right ) | 0 \rangle \, e^{i q x} \, d^4
x
\label{pop}
\end{eqnarray}
(the mass of the $b$ quark is entirely neglected throughout this paper).
The electroweak correction $\Delta \rho$ is found as$^{\cite{barb}}$

\beq
\Delta \rho= { {G_\mu} \over \sqrt{2}} \, m_t^2 \, ( P_\pm^\prime(0) -
P_0^\prime(0))~~,
\label{dif}
\eeq
where $P^\prime(q^2)=dP(q^2)/dq^2$.

Consider now a calculation of the difference of the derivatives of the
vacuum polarization operators in eq.(\ref{dif}) by the conventional Feynman
diagram technique. The one-loop graphs (no additional gluons) give
logarithmic divergence for each of the derivatives, but their difference is
finite and the result is the leading term in eq.(\ref{rho}).
Originally
the two-loop graphs, which lead to the $O(\as)$ correction in eq.(1),  were
calculated$^{\cite{dv}}$ by using the dispersion relations. Here we
concentrate on the direct calculation of these graphs by Feynman's
technique, which transparently reveals the structure of Euclidean distances
contributing to the first QCD correction. For this purpose we first consider
the integration over the quark loop, which allows to represent the result as
an integral over the gluon momentum. One can notice that starting with
the first order QCD correction the integration over the quark loop is
finite for each of the derivative terms in eq.(\ref{dif}). Also at $q^2
\approx 0$ there is no obstruction to the Wick rotation. Thus we find in
terms of integrals over the Euclidean momentum of the gluon the
expressions for the first QCD corrections to the derivatives of the vacuum
polarization operators (\ref{pop}):
\begin{eqnarray}
\delta^{(1)} P_\pm^\prime(0) &=& {1 \over {4 \pi^3}} \int
w_\pm \left ({ k \over {m_t}} \right ) \, \as
\, d(k^2) k^2 { {d\,k^2} \over {m_t^2}} ~, \nonumber \\
\delta^{(1)} P_0^\prime(0) &=& {1 \over {4 \pi^3}} \int
w_0 \left ({k \over {m_t}} \right ) \, \as
\, d(k^2) k^2 { {d\,k^2} \over {m_t^2}}
\label{popint}
\end{eqnarray}
with $d(k^2)=1/k^2$ being the transverse gluon propagator{\footnote {The
expressions (\ref{popint}) are gauge-invariant, hence it is only the
transverse gluon propagator, which is contributing.}} and the weight
functions $w_\pm$ and $w_0$, determined from the quark loop integration, are
given by
\beq
w_\pm (x)= 2\,{{ 12+16\, x^2+ 8 \, x^4+ x^6} \over {x\,\sqrt{4+x^2}}} \,
{\rm arctanh} \left ( {x \over \sqrt{4+x^2}} \right ) - 2{ {(1+x^2)^3} \over
{x^2}}\, \ln (1+x^2) + x^4 \, \ln x^2 - 4
\label{wpm}
\eeq
and

\beq
w_0(x)=12 {{2+x^2} \over {x \, (4+x^2)^{3/2}}} \, {\rm arctanh }
\left ( {x \over \sqrt{4+x^2}} \right ) - { 6 \over {4+x^2}}
\label{w0}
\eeq
Correspondingly the difference, entering eq.(\ref{dif}), can be written in
the form

\beq
P_\pm^\prime(0) - P_0^\prime(0) =  {1 \over {4 \pi^3}} \int
w \left ({ k \over {m_t}} \right ) \, \as
\, d(k^2) k^2 { {d\,k^2} \over {m_t^2}}
\label{difint}
\eeq
with the resultant weight function
\begin{eqnarray}
&& ~~~~~~~~~~~~~~~w(x)= w_\pm (x)-w_0 (x) = \nonumber \\
&& 2\, {{36 + 70\,{x^2} + 48\,{x^4} + 12\,{x^6} +
{x^8}} \over {x \, (4+x^2)^{3/2}}} \, {\rm arctanh} \left ( {x \over
\sqrt{4+x^2}} \right ) - \nonumber \\
&& 2\, { {(1+x^2)^3} \over
{x^2}}\, \ln (1+x^2) + x^4 \, \ln x^2 - 2\, {{5+2\,x^2} \over {4+x^2}}
\label{wdif}
\end{eqnarray}

At large $x$ the function $w(x)$ has the asymptotic behavior
$w(x) = 3 x^{-2} + O(x^{-4})$, so that the integral in eq.(\ref{difint}) is
logarithmically divergent. This divergence is regularized, once the gluon
propagator is regularized by any standard procedure. Here we do not need to
specify the  regularization procedure, since it can be easily noticed that
the expression (\ref{dif}) for the measurable quantity $\Delta \rho$
contains also the factor $m_t^2$ and the divergence in eq.(\ref{difint}) is
the same as in the renormalization of $m_t^2$. More specifically the $\as$
correction to $\Delta \rho$ in terms of the top quark on-shell mass is
determined by

\beq
\delta \left ( m_t^2 \, (P_\pm^\prime(0) - P_0^\prime(0)) \right ) =
m_t^2 \, \left ({2 \over m_t}\, \Sigma(m) + \delta
(P_\pm^\prime(0) - P_0^\prime(0)) \right ) ~,
\label{mr}
\eeq
where $\Sigma(\gamma \cdot p)$ is the one-loop top quark self-energy.
The on-shell value of the self-energy can be written as an integral over
the Euclidean $k^2$ of the gluon in the loop as

\beq
{2 \over m_t}\, \Sigma(m) = {1 \over {4 \pi^3}} \int
s \left ({ k \over {m_t}} \right ) \, \as
\, d(k^2) k^2 { {d\,k^2} \over {m_t^2}}
\label{sw}
\eeq
with the weight function

\beq
s(x)={{x^4+2 \,x^2-8} \over {2\, x\, \sqrt{4+x^2}}} - {{x^2} \over 2}~~.
\label{sx}
\eeq
Therefore the final expression for the first $\as$ correction to $\Delta
\rho$ (eq.(\ref{mr})) in terms of the on-shell mass $m_t$ is proportional to
the finite integral

\beq
\int_0^\infty \left ( w(x)+ s(x) \right ) d x^2 = -{{\pi^2} \over 3}-1~~,
\label{fint}
\eeq
which reproduces the known result$^{\cite{dv}}$ in eq.(\ref{rho}).

However, the structure of the integral in eq.(\ref{fint}) deserves a closer
look. Indeed, as one can see from the plots of the weight functions $s(x)$
and $w(x)$ shown in Fig.1, the integral is significantly contributed by the
region of small $x$ due to the $-2/x$ behavior of the function $s(x)$ in
that region, i.e. at $k \ll m_t$. This behavior is clearly a
consequence of the fact that the on-shell mass $m_t$ is chosen as the
parameter in the electroweak loop. As to the weight function of the
electroweak loop itself, $w(x)$, as is clearly seen from the plot, it is
completely dominated by the region $x \ge 1$, i.e. $k \ge m_t$, and thus
it displays practically no sensitivity to long distances.

The significance of the contribution of the region of small Euclidean $k$
can be quantified in this calculation by applying the BLM
criterion$^{\cite{blm}}$ for the normalization point of $\as$. The BLM
procedure amounts to replacing the bare gluon propagator
$\as \, d(k^2)= \as/k^2$ by the one with the running coupling constant:
$ \as^V(k) /k^2$, where $\as^V(k)$ is the effective coupling constant in
the potential between two infinitely heavy quarks. The running constant
$\as^V(k)$ in the Coulomb gauge corresponds to including the vacuum
polarization insertions in the propagator of the Coulomb gluon. In
particular the BLM procedure applied to the calculation in the first order
in $\as$ correctly reproduces the dependence of the coefficient of the next
term $\as^2$ on the number $n_f$ of light quark flavors, since these enter
only through the loop insertion in the gluon propagator. However, this
procedure additionally combines the $n_f$ dependence into the factor
$b_0=11-{2 \over 3} n_f$, which is the first coefficient of the QCD
$\beta$-function. In numerous examples$^{\cite{blm,b2}}$ applying this
criterion to the choice of the normalization point for the coupling constant
removes large coefficients in the subsequent term{\footnote {The only known
cases, where large numerical coefficients are not removed by this procedure
are those associated with the annihilation of heavy quarkonia in
QCD$^{\cite{blm,b2}}$, which parallels a similar behavior of the
three-photon annihilation of ortho-positronium in QED.}}.

To apply the BLM procedure to the calculation of $\Delta \rho$ we use
in the integrals in eqs.(\ref{difint}, \ref{sw}) the one-loop effective
coupling:

\beq
\as \, d(k^2) \to { 1 \over {k^2}} \, \as^V(m_t) \, \left ( 1- { {\as^V(m_t)
\, b_0} \over {2 \pi}} \ln (k/m_t) \right ) ~.
\label{run}
\eeq
Then $\as$ times the integral in eq.(\ref{fint}) is replaced by

\begin{eqnarray}
&& \as^V(m_t) \,\int_0^\infty \left ( 1- { {\as^V(m_t)
\, b_0} \over {2 \pi}} \ln (x) \right ) \,
 \left ( w(x)+ s(x) \right ) d x^2 \approx \nonumber \\
&& -\as^V(m_t) \,
 \left ( {{\pi^2} \over 3}+1 \right ) \, \left ( 1+ 1.034 \, { {\as^V(m_t)
\, b_0} \over {2 \pi}} \right ) \approx  - \as^V(0.355\,m_t) \,
 \left ( {{\pi^2} \over 3}+1 \right ) ~~,
\label{normv}
\end{eqnarray}
which fixes the normalization point in terms of $\as^V$
(the integral with the factor $\ln (x)$ was calculated numerically.)
Furthermore, the normalization of the effective coupling
$\as^V(k)$ is simply related$^{\cite{blm}}$ to that of the $\as^\msb$:
$\as^V(k) = \as^\msb(e^{-5/6}\,k) \approx \as^\msb(0.435\,k)$. Therefore
from eq.(\ref{normv}) we
find that within the BLM scheme the effective coupling, entering the first
QCD correction in eq.(\ref{rho}) is $\as^\msb(0.154 \, m_t)$. It is clear,
however, that to completely quantify the magnitude of the QCD correction to
$\Delta \rho$ a full three-loop calculation of the terms $\as^2$ is needed.

It is clear that a similar calculation, operating only with Euclidean-space
integrals over the momenta of the gluons can be in principle performed in
higher orders of the QCD perturbation theory, thus making it free from the
long-distance uncertainties. Exactly this point was discussed at length in
connection with the QCD sum rules in the papers \cite{sr} and also in a
later review \cite{rry}. Here we would like to point out a specific loophole
in the reasoning of the recent papers$^{\cite{ks1,ghv,ks2}}$, which state
that through the dispersion relations the contribution of the near-threshold
region to $P^\prime (0)$ makes the latter quantity sensitive to the
long-distance dynamics. We also disagree with the argumentation of
Ref.\cite{ynd}, where it is argued that the threshold effects are {\it
numerically } small. We insist here that the long-distance effects are small
{\it parametrically}, i.e. suppressed by powers of $\Lambda_{QCD}/m_t$ at
the non-perturbative level, and are absent altogether in any finite order of
perturbation theory in QCD.

The reasoning in those papers is as follows. The difference of the
derivatives of the vacuum polarization operators in eq.(\ref{dif}) can be
written in the form of the dispersion integral

\beq
P_\pm^\prime(0) - P_0^\prime(0)={1 \over \pi} \int {{\rho_\pm (s)-
\rho_0(s)} \over {s^2}} \, ds ~~
\label{disp}
\eeq
where $\rho_\pm(s)$ and $\rho_0(s)$ are respectively the spectral
densities of the operators $(\tb \, (1-\gamma_5) \, b)$ and
$i\,\sqrt{2} \, (\tb \gamma_5 t)$, and the integral is running over all
values of $s$, where the spectral densities are non-zero.
Consider now the region of $s$ near the $t \tb$ threshold, where the
integral is contributed by the $t \tb$ resonances and the very beginning of
the continuum, strongly distorted by long-distance interactions. Within the
perturbation theory the exchange of Coulomb gluons between the quark and the
antiquark with a small velocity $v=\sqrt{1-4\,m_t^2/s}$ makes the
QCD effects depend on the parameter $\as/v$ rather than $\as$. Therefore
at $v$ of the order of or less than $\as$ these effects should be summed up.
The summation amounts to using the well known solution of the Coulomb
problem, and the net effect reduces to multiplying the bare spectral density
$\rho_0^{(0)}$ by the Coulomb factor

\beq
F_c={{4 \pi \, \as /3 v} \over {1-\exp (-4 \pi \, \as /3 v)}}~.
\label{fc}
\eeq
At $v \sim \as$ the spectral density is of order of $\as$ and the
size of this region of integration in eq.(\ref{disp}) is
$\Delta  s \sim 4m_t^2\, \Delta v^2 \sim m_t^2 \as^2$. Therefore the
contribution of the `Coulomb' region above the threshold in eq.(\ref{disp})
is of order $\as^3$, which is the same as that of the under-the-threshold
resonances. The point of the papers \cite{ks1,ghv,ks2} is that the $\as$ in
this effect is normalized at long distances: $\as  (m_t \, v \sim m_t \as)$.
Due to favorable numerical factors of $\pi$ this $O(\as^3)$ effect is stated
to be a sizeable fraction of the $\as$ term in eq.(\ref{rho}). To quantify
this statement the dispersion integral over the near-threshold region is
calculated$^{\cite{ghv,ks2}}$ with the factor $F_c$ in eq.(\ref{fc}) in
which the running of the coupling constant is parametrized as $\as(m_t
v)${\footnote {More precisely, in Ref.\cite{ghv} a relativistic
parametrization is used, which however, does not change the main point of
the argument.}}.

The loophole in this argument is that in terms of the running constant $\as$
in eq.(\ref{fc}) the normalization point {\it is not} given by $m_t \, v$.
Rather, the proper normalization point is related to $m_t \, v$ by a
function $f(\as/v)$: $\as (m_t \, v \, f(\as/v))$. This phenomenon is
clearly seen in the QED calculation$^{\cite{sv}}$ of the excitation curve of
the $\tau^+ \, \tau^-$ at the threshold, including the Uehling-Serber
running of the QED constant $\alpha$.  Namely, in the region $v \sim \alpha$
the formula, obtained by the simple substitution of $\alpha$ by
$\alpha(m_\tau \, v)$ significantly deviates from the exact result.  The
contribution of the sum over resonances and of the integral over the
continuum in eq.(\ref{disp}) in higher orders in $\as$ contains delicate
cancellations$^{\cite{sr,ynd}}$:  starting from order $\as^3$ the integral
over the continuum partly compensates the contribution of
resonances{\footnote {The objection$^{\cite{ks2}}$ that these terms in the
integral over the continuum can not be negative, since the quark and the
antiquark are attracting each other, is obviously erroneous: these are small
negative corrections of order $\as^3$ to the positive contribution in the
orders $\as$ and $\as^2$.}}.  Moreover, in calculation of the dispersion
integral the function $F_c$ cannot be expanded in powers of $\as$, since
this expansion does not converge at $v < \as$ (within such expansion the
integrals of individual terms would diverge at $v = 0$ starting from the
order $(\as/v)^3$).  The result of the integration however can be perfectly
expanded as a series in $\as\,^{\cite{sr}}$. Thus an approximate
parametrization is certainly prone to giving misleading results by
destroying the correct structure of the spectral density at small $v$.
Therefore we conclude that clarifying this point by calculating the spectral
density near the threshold is a far more complicated problem, than the
initial one of calculating the vacuum polarization far below the threshold.
On the other hand, those detailed calculations of the near-threshold region
are not needed, since, as discussed before, the $O(m_t^2)$ electroweak
corrections can be calculated by the Feynman diagrams entirely in the
Euclidean space, in which at no step the long-distance uncertainty of the
QCD dynamics shows up at the perturbative level. In other words, though the
integrand in eq.(\ref{disp}) is poorly calculable near the $t \tb$
threshold, the entire integral is well calculable within the short-distance
QCD.

As to the non-perturbative QCD effects, these can be understood by adapting
the results of the discussion of the charmonium sum rules (sections 7.3 --
7.6 of the second paper in Ref.\cite{sr}). The result is that any finite
distortion of the quark-antiquark interaction at a finite distance $r_0 \gg
m_t^{-1}$ produces only an effect on the $t \tb$ vacuum polarization at $q^2
\approx 0$, which is suppressed by $\exp (-2\, m_t \, r_0)$. For instance,
one can cut off the Coulomb interaction at a radius $r_0 \ll (m_t \,
\alpha)^{-1}$ (but still $r_0 \gg m_t^{-1}$), so that the Coulomb-like bound
states disappear, and the actual spectral density $\rho_0$ would look
nothing like that determined by the factor $F_c$ in eq.(\ref{fc}). Still, up
to the exponentially suppressed terms the vacuum polarization at $q^2
\approx 0$ in this situation would be given by the dispersion integral in
eq.(\ref{disp}) with the {\it perturbative} spectral density, i.e. the one
containing Coulomb-like poles, and the factor $F_c$ above the threshold.
Fully appreciating the non-trivial character of this phenomenon, we
point out that this is a direct consequence of the analyticity
of quantum amplitudes.
The only way, in which the long-distance effects give a contribution to
the vacuum polarization at $q^2 \approx 0$ is through the `tail' of the
long-distance effects at distances of order $m_t^{-1}$. In the potential
models with a power-like non-perturbative potential of the form
$V(r) = a \, r^n$, the effect is proportional$^{\cite{sr}}$ to $a \,
m_t^{-(n+1)}$ (which is the action $\int V(r) \, dt$ at distances
$r \sim m_t^{-1}$ over the time $t \sim m_t^{-1}$). To evaluate this
particular effect in QCD there is however no need of invoking
model potentials, and the leading effect is calculable in terms of
the vacuum gluon condensate and its relative magnitude is
given by $\langle 0 | \pi \as G_{\mu
\nu}^a G_{\mu \nu}^a | 0 \rangle /m_t^4 \sim 10^{-10}$. This would be the
only non-perturbative contribution to the $O(m_t^2)$ corrections, if these
corrections were expressed in terms of the top mass, normalized at short
distances. However when the corrections are expressed through the on-shell
mass of top, the relative non-perturbative contribution is that in the
on-shell mass, i.e. $O(\Lambda_{QCD}/m_t)$.

We are thankful to L.B. Okun for stimulating discussions.
This work was supported, in part, by the DOE grant DOE-AC02-83ER40105.

{\large \bf Figure Caption} \\[0.15in]

{\bf Figure 1.} The weight functions $s(x)$ (left) and $w(x)$ (right) vs.
$x=k/m_t$,
in the integrals over the Euclidean gluon momentum in the top quark
self-energy (eq.(\ref{sw})) and in the heavy quark loop (eq.(\ref{difint})).
Note the strongly different scale of the vertical axis in the plots.

\end{document}